\documentclass[10pt,conference]{IEEEtran}

\usepackage{cite}
\usepackage[linesnumbered,ruled]{algorithm2e}

\usepackage{balance}
\usepackage{threeparttable}
\usepackage{graphicx}
\usepackage{textcomp}
\usepackage{xcolor}
\usepackage{comment}
\usepackage{booktabs}
\usepackage{multirow}
\usepackage{xspace}
\usepackage{tcolorbox}

\usepackage{array}

\usepackage{subcaption}

\usepackage{amsmath}
\usepackage{amssymb}
\usepackage{mathtools}
\usepackage{amsthm}

\newcommand{\PreserveBackslash}[1]{\let\temp=\\#1\let\\=\temp}
\newcolumntype{C}[1]{>{\PreserveBackslash\centering}p{#1}}
\newcolumntype{R}[1]{>{\PreserveBackslash\raggedleft}p{#1}}
\newcolumntype{L}[1]{>{\PreserveBackslash\raggedright}p{#1}}

\newif\ifshowtodos
\showtodostrue

\def\BibTeX{{\rm B\kern-.05em{\sc i\kern-.025em b}\kern-.08em
    T\kern-.1667em\lower.7ex\hbox{E}\kern-.125emX}}
    
\newcommand{\pa}[1]{\noindent\textbf{#1}}

\newcommand{\toolS}{\textit{SORBET}\space}
\newcommand{\tool}{\textit{SORBET}}

\newcommand{\RQOne}{How robust are LiDAR-based 3D obstacle detection models against specification-based perturbations?}
\newcommand{\RQTwo}{What are the cascading impacts on ADSs caused by the perturbations of the LiDAR's PCD input?}
\newcommand{\RQThree}{Can retraining using PCD with perturbations improve the robustness of ADSs?}
\newcommand{\rqboxc}[1]{\begin{tcolorbox}[left=4pt,right=4pt,top=4pt,bottom=4pt,colback=gray!5,colframe=gray!40!black,before skip=4pt,after skip=4pt]#1\end{tcolorbox}}

\AtBeginDocument{%
  \providecommand\BibTeX{{%
    Bib\TeX}}}

\begin{document}

\title{On the Robustness Evaluation of 3D Obstacle Detection Against Specifications in Autonomous Driving}

\author{\IEEEauthorblockN{Tri Minh-Triet Pham}
\IEEEauthorblockA{\textit{O-RISA Lab} \\
\textit{Concordia University}\\
Montreal, Quebec, Canada \\
p\_triet@encs.concordia.ca}
\and
\IEEEauthorblockN{Bo Yang}
\IEEEauthorblockA{\textit{O-RISA Lab} \\
\textit{Concordia University}\\
Montreal, Quebec, Canada \\
b\_yang96@encs.concordia.ca}
\and
\IEEEauthorblockN{Jinqiu Yang}
\IEEEauthorblockA{\textit{O-RISA Lab} \\
\textit{Concordia University}\\
Montreal, Quebec, Canada \\
jinqiu.yang@concordia.ca}
}


\maketitle

\begin{abstract}

Autonomous driving systems (ADSs) rely on real-time sensor data, such as cameras and LiDARs, for time-critical decisions using deep neural networks. The accuracy of these decisions is crucial for the widespread adoption of ADSs, as errors can have serious consequences. 3D obstacle detection, in particular, is sensitive to point cloud data (PCD) noise from various sources. However, the robustness of current 3D obstacle detection models against specification-based perturbations remains unevaluated. These perturbations are derived from the specification of LiDAR sensors and previous research on LiDAR's ability to capture objects of different colors and materials. They can manifest as very subtle sensor-based noises or obstacle-specific perturbations.
Hence, we propose \tool, a framework that tests the robustness of 3D obstacle detection models in ADS against such perturbations to the PCD to evaluate their robustness.
We applied \tool{} to evaluate the robustness of five classic 3D obstacle detection models, including one from an industry-grade Level 4 ADS (Baidu's Apollo). Furthermore, we studied how the deviated obstacle detection results would propagate and negatively impact trajectory prediction.
Our evaluation emphasizes the importance of testing 3D obstacle detection against specification-based perturbations. We find that even very subtle changes in the PCD (i.e., removing two points) may introduce a non-trivial decrease in the detection performance. 
Furthermore, such a negative impact will further propagate to other modules and endanger the safety of the ADS. 
\end{abstract}


\section{Introduction}
Automated driving systems (ADSs) have gained significant attention in both research and industry, driven by their substantial commercial potential~\cite{Tang2022ASO, waymo, driverless-taxis, driverless-grocery} and the high costs associated with their errors~\cite{kpmg_driverless_car, Cui2019ARO}.
In ADSs, perception is safety-critical.
Perception uses a combination of algorithms and deep neural networks (DNNs) to process sensor scans, e.g., camera and LiDAR (Light Detection and Ranging), allowing interpretation of the environment in real-time. One crucial functionality of perception is obstacle detection, which determines the type, size, and location of nearby obstacles in the surroundings.
Obstacle detection processes data from different sensors such as cameras, LiDARs, and radars, and combines them to detect and locate obstacles.
Obstacles include vehicles, pedestrians, cyclists, static roadside objects, etc., which are determined by post-processing algorithms.
These obstacles are crucial for subsequent tasks such as trajectory prediction and path planning for appropriate interaction.
The key component of this pipeline is 3D obstacle detection, which detects obstacles from LiDAR input. 
LiDAR sensors are often considered the main sensor \cite{apollo_main_sensor_sample} in obstacle detection due to their ability to accurately detect and map all obstacles in a 360\textdegree~view around the ADS under low light conditions.
However, LiDAR sensors are imperfect and suffer subtle inaccuracies. For example, three popular LiDAR sensors: Velodyne HDL-32E \cite{nuscenes}, Velodyne HDL-64E \cite{lidar_manuals, Geiger2012CVPR}, and Ouster OS2 \cite{lidar_manuals} have distance accuracies of up to 2cm, 5cm, and 10cm, respectively.
Factors affecting the accuracy include, but are not limited to, the difference in position, angle of capture, etc., of both the sensor and the obstacle between the start and the end of each LiDAR sweep, sensor calibration, and reflectivity of the surfaces.
In addition, since LiDAR sensors utilize lasers for scanning, color (and the material that produced the color), weather, and lighting conditions of the environment can significantly change how much light the sensor receives.
Such factors augment the point cloud data (PCD) captured by LiDAR, posing challenges to the robustness of the underlying DNNs in ADS's 3D obstacle detection. 

Despite tremendous works on robustness evaluation for 3D object detection against against adversarial attacks \cite{Xu2021AdversarialAA, adversarial_lidar_attack, invisibleobject} and corruptions \cite{Tang2022ASO, leimalidarrobustnessbenchmark, Dong2023BenchmarkingRO, Sun2022BenchmarkingRO}, there exists a research gap in the robustness evaluation of 3D obstacle detection against realistic specification-based perturbations based on (1) official LiDAR sensor specifications and (2) prior studies showing obstacle-specific characteristics, e.g., color and material, influence the PCD captured by LiDAR.
Existing synthetic corruptions are commonly based on extreme weather conditions \cite{kilic2021lidar}, sensor vibrations and noise \cite{Ma2012AnalysisOP}, lighting, and surface reflections. However, the connection between these simulated corruptions and their real-world causes is often unclear, particularly in terms of how specific parameter values are chosen. Most studies assume that such sensor corruptions are already reflected in the PCD, leading to the use of theoretical noise models to simulate their effects. 
As it remains unclear which types of errors are being referenced or which sensor characteristics are responsible for the observed perturbations, existing works \cite{leimalidarrobustnessbenchmark, Dong2023BenchmarkingRO} use a wide range of synthetic corruption values.
In many cases, the tested perturbation ranges significantly exceed, and often barely overlap with, the subtle levels of sensor inaccuracies observed in real-world scenarios.
Moreover, current robustness benchmarks for 3D obstacle detection focus on scene-wide effects from environmental conditions (e.g., weather, lighting) or significant changes to the objects, e.g., occlusion. This overlooks how object-specific properties, such as color, material, or texture, can alter PCD even under normal sensor operation \cite{Sequeira2021ColorPerturbation}. These factors remain underexplored, and when included, are often grouped with other perturbations without clear analysis, limiting insight into their individual impact.
For example, we applied a z-test to compare perturbed PCDs by normal range inaccuracy (one of our perturbations) and Gaussian noise~\cite{Dong2023BenchmarkingRO}, which are similar as both apply Gaussian noise to the PCD, and found a significant difference (z = -62.0, p $<$ 0.001).

Lastly, the impact of these perturbations on downstream ADS tasks remains underexplored. While the ideal scenario assumes that each system component functions with minimal error, current literature has yet to investigate the consequences of misdetected or misrepresented obstacles. Understanding how such errors propagate through the pipeline is crucial for evaluating the real-world robustness and safety of ADSs.
Incorporating these often-overlooked factors into the testing space can provide a more comprehensive and realistic assessment of model robustness under diverse real-world conditions.
Since ADSs are complex, multi-module systems that are inherently safety-critical, enhancing their testing processes has implications that extend to other LiDAR-based autonomous systems, such as robots and drones, with applications in agriculture, medicine, aviation, and maritime domains. These systems represent a significant and growing class of software currently in use.


To address these concerns, in this paper, we propose \tool, a framework to evaluate the robustness of 3D obstacle detection against real-world specification-based perturbations.
\tool{} injects specification-derived perturbations into PCDs to evaluate the robustness of representative 3D obstacle detection models and measures the resulting deviation in detection.
Then we evaluate the propagation of such perturbations on subsequent decision-making modules (e.g., prediction) in ADS via the measured deviations.
We apply \tool{} to evaluate the robustness of five representative obstacle detection models, including the model in Apollo, an industry-grade ADS from Baidu. Furthermore, we performed a simulation-based study to comprehensively assess the impact of deviated obstacle detection results on both models from academia, such as Trajectron++, and an industry-grade ADS Apollo.
Our evaluation shows that the perturbations in \tool{} can decrease the number of obstacles detected by 23.5\%. For the remaining detected obstacles, their positions can be shifted by up to 2.3m. These deviations can cascade and shift the predicted trajectory of these obstacles by up to 9.77m.


Our work offers the following contributions:
\begin{enumerate}
    \item \textbf{Specification-derived perturbations.} We propose a set of distinctive perturbations grounded in LiDAR specification and previous experiments, several of which are entirely new, while others differ significantly from prior works by using specification-derived error values.
    Our perturbations naturally complement the search space of existing work and enhance the realism of their tests.
    \item 
    \textbf{Evaluation of propagated impacts}. We go beyond standalone detection and assess how detection deviations propagate to trajectory prediction, which has not been addressed in prior studies.
    \item 
    \textbf{Adaptability through retraining.} We investigate the effects of retraining detection models on our perturbations to evaluate their adaptability to specification-driven perturbations.
\end{enumerate}

Together, these contributions establish \tool{} as the first framework to systematically evaluate specification-driven perturbations with their cascading impacts on ADS performance.

\pa{Paper organization.} The remainder of the paper is organized
as follows. Section~\ref{sec:background} presents the background and related work.
Section \ref{sec:method} introduces \tool, an automated framework to evaluate 3D obstacle detection and further assess its impact on trajectory prediction.
Section~\ref{sec:setup} introduces the setup of our experiments.
Section~\ref{sec:results} introduces the evaluation of \tool.
Section~\ref{sec:threats} discusses the threats to the validity of this work.
Section \ref{sec:discussion_future} discusses our plan for future work, and finally Section~\ref{sec:conclusion} concludes the paper.

\section{Background and Related Work}
\label{sec:background}
In this section, we first provide a brief overview of ADS testing at a high level, and then discuss LiDAR-based 3D obstacle detection and trajectory prediction in detail, as these are the primary components targeted in this work.
While none of the previous works test the robustness of 3D obstacle detection to specification-derived perturbations and its impact on trajectory prediction, several have tested them as independent modules. 

\subsection{Testing and attacking ADS overview}
Testing of autonomous driving systems (ADS) has become a major research focus due to the safety-critical consequences of failures, spanning evaluations from individual models to end-to-end system testing~\cite{10.1145/3579642} and covering a wide range of tasks, including traffic-light recognition~\cite{10588456,autosec:2021:kanglan:roiattack}, localization~\cite{yang2025robustnesslidarbasedposeestimation,shen2020drift}, obstacle detection and trajectory prediction (discussed in detail in the following sections), planning~\cite{wan2022afraiddrivesystematicdiscovery}, and system-level behavior~\cite{10064002}.

\begin{figure*}[ht]
\centerline{\includegraphics[width=0.8\textwidth,keepaspectratio]{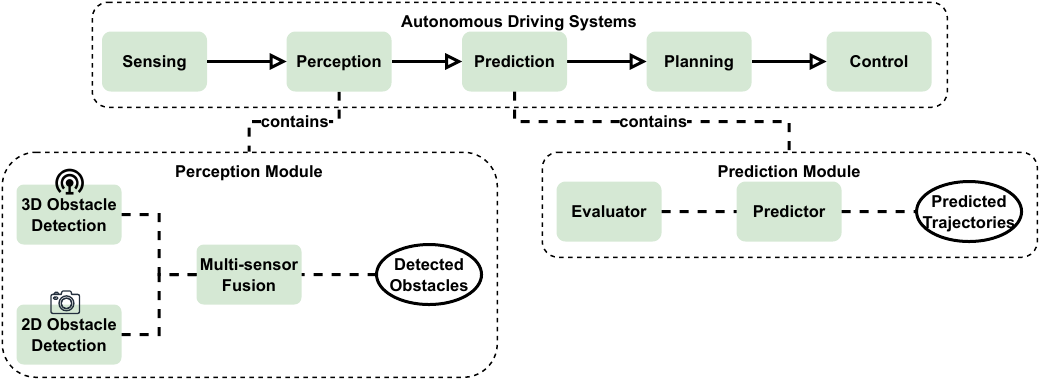}}
\caption{An Overview of a typical ADS}
\label{fig:typical_ads}
\end{figure*}
\subsection{Testing and attacking 3D obstacle detection}
3D obstacle detection relies on LiDAR to measure accurate distances and detect obstacles under all lighting conditions, providing full 360\textdegree{} coverage. At fixed intervals, the LiDAR sensors scan the environment and store it as PCD, a set of data points of the 3D position and the intensity, representing the reflected light from the environment. 
The PCD is then processed by 3D obstacle detection systems, which output 3D bounding boxes indicating the locations and distances of detected obstacles.

\pa{Testing 3D obstacle detection.} 3D obstacle detection has been subjected to significant testing efforts, including metamorphic testing \cite{metamorphic_lidar}, fuzzing testing~\cite{pham2024perceptionguidedfuzzingsimulatedscenariobased} and 
weather and sensor-based PCD corruptions testing \cite{yu2022benchmarking, Gao2023BenchmarkingRO, leimalidarrobustnessbenchmark, LiRtest, Dong2023BenchmarkingRO}.
Similar weather- and sensor-based tests have also been conducted on LiDAR-based general object detectors \cite{Albreiki2022OnTR, Sun2022BenchmarkingRO}, which are limited to recognizing a wide range of objects solely from point clouds of the objects themselves without considering the environment and distances.

\pa{Attacking 3D obstacle detection.} Beyond testing, significant efforts have focused on attacking this module via the PCD through various methods to expose its vulnerabilities. This includes attacks on regions \cite{Xu2021AdversarialAA, ZHENG2023109825,adversarial_locations, Zhu2021AdversarialAA} or obstacles in the PCD. Obstacle-based attacks can target existing obstacles using guided perturbations \cite{lidar_occulsion, Yang2021RobustRP, WANG202127, gan_attack,liu_multiview}, shifting \cite{Hau2021ObjectRA}, or occlusion patterns \cite{lidar_occulsion}. On the other hand, it can generate new obstacles via spoofing points \cite{adversarial_lidar_attack} or generating obstacles with adversarial shapes \cite{invisibleobject, Illusion_and_Dazzle, pmlr-v164-tu22a, Tu_2020_CVPR, Abdelfattah}. 

In this work, we test LiDAR obstacle detection assuming normal, non-adversarial conditions without external interference. There are several key differences between our work and related works \cite{Sun2022BenchmarkingRO, Dong2023BenchmarkingRO, LiRtest}.
First, while prior works benchmark a wide range of noise for diverse hypothetical scenarios (indicating hardware/software failure or interference), \tool{} focuses on realistic specification-based perturbations (that can happen when the instrument is operating under normal conditions). These two approaches result in minimal overlap. Our maximum combined 3D shift is 2cm, significantly smaller than the 5cm per-axis shift used in previous studies. Moreover, prior work perturbs each dimension independently, whereas \tool{} applies perturbations in the combined 3D space, leading to a statistically distinct distribution. For example, comparing perturbed PCDs from range inaccuracy (\tool) and Gaussian noise \cite{Dong2023BenchmarkingRO} using the z-test shows a significant difference (z = -62.0, p $<$ 0.001).
Also, some data \cite{Sun2022BenchmarkingRO} only contains the points of an object and cannot be used to evaluate the 3D obstacle detector.
Second, we introduce localized perturbations based on values from prior work \cite{Sequeira2021ColorPerturbation}, targeting real-world phenomena such as point cloud variations caused by object color differences, e.g., different colors of the same car model.
Third, we assess the impact of these perturbations on all 3D obstacle detections, including those that fall below detection thresholds, to provide a more comprehensive analysis.
Lastly, we evaluate how these detection deviations propagate to downstream ADS modules, offering deeper insight into the broader effects of the perturbations.
Overall, our perturbations are realistic, naturally extend the test space of 3D detectors, and complement existing research.

\subsection{Trajectory prediction tests and attacks}
Trajectory prediction is a core component in ADSs, which predicts the trajectory of the detected obstacles in the next few seconds based on their historical position, heading, velocity, etc. 
\pa{In Apollo}, the standalone prediction module \cite{apollo} predicts the position of all detected obstacles using their historical positions, headings, velocities, etc.
The module comprises four components, i.e., container, scenario analyzer, evaluator, and predictor.
Each combination of obstacle type and scenario has its own trajectory prediction model~\cite{pengzi}.

\pa{Attacking trajectory Prediction.} Several works have attacked this module since its output affects the reaction of the ego car. This includes adversarial attacks on the vehicle trajectory history \cite{Zhang_2022_CVPR, advsim, SAADATNEJAD2022103705}, adversarial patches to tracking \cite{attack_tracking}, and spoofing the ego car's trajectory \cite{li2021fooling}.
We are not proposing an attack on the trajectory prediction module.

\pa{Testing trajectory prediction.}
While significant effort is invested in attacking trajectory prediction systems, few test the module.
Current works include metamorphic testing to detect unavoidable collisions \cite{metamorphic_prediction} and auto-encoder-based adversarial training to enhance the robustness \cite{Jiao2022SemisupervisedSA}.
Different from existing works, in this paper, we seek to test the robustness of trajectory prediction to specification-derived perturbations. Our experiments seek to quantify the impact of the propagation of such perturbations on trajectory prediction.

\section{\tool: An Automated Framework for Evaluating the Robustness of ADSs Against Specification-Based Perturbations}
\label{sec:method}
\begin{figure*}[htbp]
\centerline{\includegraphics[width=0.8\textwidth,keepaspectratio]{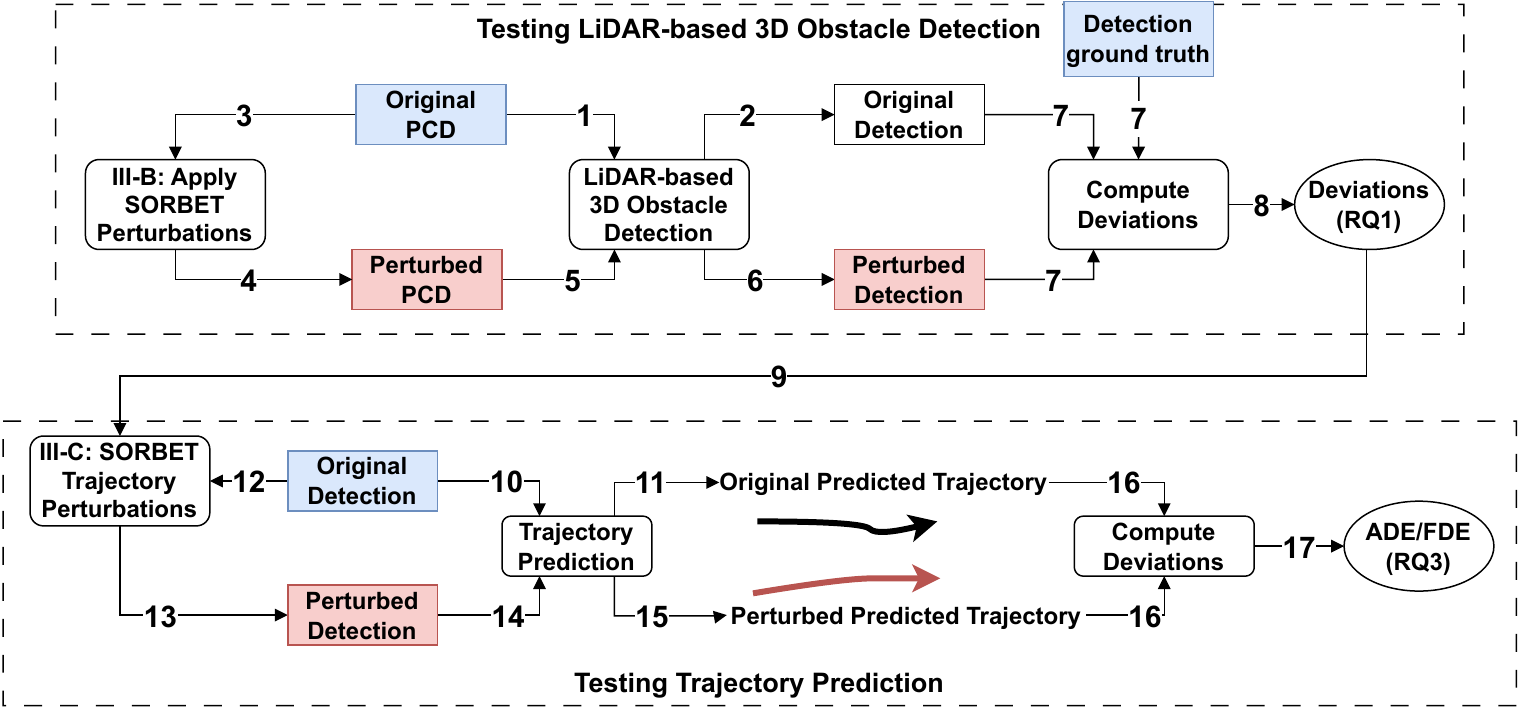}}
\caption{Overview of the SORBET framework with numbered steps indicating the dataflow order.
}
\label{fig:implementation}
\end{figure*}
\pa{Overview of \tool.} In this section, we present the design of \toolS as (1) a standalone framework testing the robustness of 3D obstacle detection and the propagation of the resulting deviation on trajectory prediction; and (2) an integration to Apollo (an industry-grade multi-module ADS) in a simulator environment.
Figure~\ref{fig:implementation} illustrates the major components of \tool: applying perturbations to the PCD, executing the 3D obstacle detection models, measuring the resulting deviated obstacles, modeling such deviations to perturb the obstacle detection history, executing the trajectory prediction models, and measuring the resulting deviated predicted trajectory.

\subsection{Designing \tool{} PCD perturbations from specifications}
\label{sec:perturbations}

\begin{table}[t]
\caption{Perturbations included in \tool}
\centering
\resizebox{\columnwidth}{!}{
\begin{tabular}{l | l | l} 
    \toprule
    \textbf{Perturbation}              & \textbf{Variation} & \textbf{Distribution} \\
    \midrule
    (1) Range Inaccuracy (RI)
                    & global, local, direction
                                    & uniform, Gaussian, Laplacian\\
    (2) False Positive  & global, local
                                            &  uniform\\
    (3) Reflectivity      
               & increase, decrease   
                                            & uniform\\
   (4) Distance Amplified RI
               & local 
                                            & uniform               \\
    \bottomrule
\end{tabular}
}
\label{tab:perturbation_types}
\end{table}


In total, we include four types of perturbations to PCD. Table \ref{tab:perturbation_types} summarizes the perturbation types designed for \tool, including their application scope and, where applicable, the random distributions used for noise or point selection. 
The first three perturbations are derived from specification documents of popular LiDAR sensors \cite{lidar_manuals}. They represent subtle, realistic inaccuracies inherent to the sensors, causing slight fluctuations in the PCD even when the scene remains unchanged. Although these fluctuations are small, prior work has shown that even minor changes in PCD can cause obstacle detection models to malfunction \cite{adversarial_lidar_attack}.
In addition, prior studies \cite{Sequeira2021ColorPerturbation} demonstrate that PCD can vary depending on object-specific characteristics such as color and material, even when sensors operate normally without interference. Building on this observation, we introduce a fourth type of perturbation to account for these effects, bringing the total to four.
Including these perturbations naturally expands the test space of PCD datasets and complements previous methods \cite{LiRtest, Dong2023BenchmarkingRO} that focus on robustness under extreme weather conditions (e.g., fog, rain) or external attacks (e.g., lasers).
We describe each of these perturbation types in detail below.

\begin{figure}[ht]
\centerline{\includegraphics[width=\columnwidth, keepaspectratio]{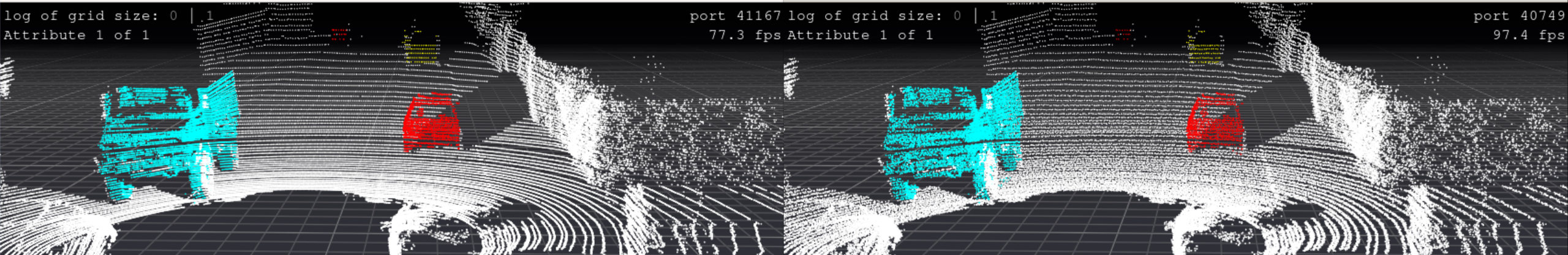}}
\caption{\textbf{Left}: Original PCD from round 19 of KITTI. \textbf{Right}: perturbed PCD using \textbf{global distance inaccuracy} where each point is moved less than 2cm away from its original position.}
\label{fig:fig_perturb_comparison}
\end{figure}

\pa{(1) Range inaccuracy.} 
Scans from LiDAR sensors can contain very subtle inaccuracies caused by imperfect calibration, angles, and viewpoints of the LiDAR hardware, as well as latencies and processing errors of LiDAR software. 
The specifications of popular LiDAR sensors~\cite{lidar_manuals, Geiger2012CVPR, nuscenes} show that the inaccuracy of the LiDAR sensors ranges from 2cm to 10cm. 
Hence, we design this perturbation to replicate range inaccuracy by shifting the xyz-coordinates of selected points in the PCD.
For our experiment, we limit the maximum perturbation of the xyz dimensions combined is $\pm$2cm.
We use the smallest inaccuracy specified, since any observed deviation would be more pronounced at higher values. In comparison, related works \cite{LiRtest, Dong2023BenchmarkingRO, Sun2022BenchmarkingRO} apply a minimum shift of 2cm in a single dimension, which is significantly larger than our maximum shift, indicating minimal overlap.
Since the specifications define only the magnitude of perturbations and not their direction, we apply them across different scopes and distributions. Noise is randomly generated using uniform, Gaussian, or Laplacian distributions (commonly used to simulate environmental noise \cite{Dosovitskiy2017CARLAAO,airsim,9967197}), while ensuring the total 3D displacement remains within the defined threshold.
We apply the perturbations with three variations: (1) global, (2) local, and (3) directional, where directional range inaccuracy has similar noise and scope as local range inaccuracy, but the perturbations have the same direction, e.g., moving all the selected points to positive x by 2cm. 
Directional perturbations replicate range inaccuracies caused by LiDAR hardware issues (e.g., calibration and view angles). The intuition behind this if the perturbation is concentrated along a direction, there is an increased likelihood that it can cause more significant deviation from the obstacle detection results.
This perturbation is applied in six directions: $\pm xyz$. 
To summarize, assume $p$ as a 3D point in PCD, we generate $\tilde{p}$ (the point with perturbation) as follows:
\begin{equation}
\tilde{p} = p + \delta, \quad \text{where } \|\boldsymbol{\delta}\|_2 \leq \epsilon
\end{equation}
where $\delta=(\delta_x, \delta_y, \delta_z)$ and $\epsilon \leq 2cm$.
An example of the resulting perturbation is shown in Figure \ref{fig:fig_perturb_comparison}.


\pa{(2) False positives.} For every point in the PCD, there is a chance that it is a false positive, thus showing a point that is not supposed to be there. We represent this phenomenon by randomly removing one point out of 10,000 points, as shown in the manual~\cite{lidar_manuals}, either from the entire PCD (global) or from an obstacle (local). Removing so few points remains unevaluated in related works.

\pa{(3) Reflectivity perturbation.}
Lighting conditions and obstacle-specific materials and colors change the absorption rate of the obstacle's surface, affecting its reflectivity and altering the intensity of the return laser captured by LiDAR sensors. 
This modifies the obstacle's point cloud representation in LiDAR scans, impacting the 3D detection results. Previous work shows that the number of points representing an obstacle in a scan varies with its color \cite{Sequeira2021ColorPerturbation}. 

Inspired by this, we apply color-based corruptions as follows: 
(1) Decreased Reflectivity: We assume that changes in the material, color, lighting, etc., of the obstacles led to a decrease in reflectivity, resulting in a reduction of 60\% of the points corresponding to those obstacles. In terms of color, this is roughly equivalent to a shift from white to black, assuming all other factors remain constant. To implement this, we randomly select and remove 60\% of the points representing each labeled obstacle from the PCD.
(2) Increased Reflectivity: We assume that the opposite has occurred, leading to an increase in point density. To implement this, we sample 67\% of the points in each obstacle and add them back to these obstacles with up to 2cm shifts in all directions, which is approximately equivalent to a color change from blue to white. The noise is chosen to increase areas where points are anticipated, i.e., no unexpected points at car windows.
Unlike related work \cite{Dong2023BenchmarkingRO}, which applies arbitrary downsampling without clear justification, we evaluate both upsampling and downsampling using point percentages informed by empirical data. These percentages reflect realistic point cloud variations caused by object color changes.

\pa{(4) Distance-amplified range inaccuracy.} 
Photonic effects such as light diffusion on complex surfaces significantly influence how LiDAR sensors receive reflected light. Surface properties play a key role: diffuse materials scatter light broadly, while retroreflective surfaces return it more directly. These variations cause range inaccuracies that depend on the obstacle's distance, angle, and surface texture relative to the ego vehicle. Unlike uniform range errors, these photonic-induced distortions affect only a subset of obstacles in the point cloud, making them harder to model and correct. For example, the OS2 sensor has a range inaccuracy of 3cm for Lambertian targets vs. 10cm for retroreflectors \cite{lidar_manuals}, showing that different parts of the same obstacle reflect light differently. Since these phenomena can cause different effects, following this perturbation, we follow the imprecision of the OS2 sensor with 10\% Lambertian reflectivity. In this case, the range inaccuracy of the obstacle changes depending on its distance $d$ to the ego car. We apply a corresponding distance-amplified shift $f(d)$ to the obstacles' point cloud.
\begin{equation}
f(d) = \pm 2.5cm \cdot \mathbf{1}_{d \leq 30} \pm 4cm \cdot \mathbf{1}_{30 < d \leq 60} \pm 8cm \cdot \mathbf{1}_{d > 60}
\end{equation}
Light-diffusion-based perturbations remain unevaluated in previous works.

\subsection{Applying \tool{} perturbations to test 3D obstacle detection}
\label{sec:local_perturb_method}
\toolS generates new PCD (with perturbations) from PCD datasets, e.g., KITTI \cite{Geiger2012CVPR}. 
In each iteration, \toolS applies one perturbation to produce one perturbed PCD, which enables individual evaluation of each perturbation.
To generate perturbations, first, \toolS reads the raw PCDs as matrices of coordinates ($O$).
For each perturbation, \toolS generates a perturbation matrix ($P$) with the same shape as the original PCD. Then, we apply the perturbation and save it as the perturbed PCD ($PCD_{pert.}$). We use these $PCD_{pert.}$ as the models' input and evaluate the resulting deviations.
\begin{equation}
    PCD_{pert.} = 
\begin{bmatrix}
o_{11} & \cdots & o_{1n} \\
\vdots & \ddots & \vdots \\
o_{n1} & \cdots & o_{nn}
\end{bmatrix}
+
\begin{bmatrix}
p_{11} & \cdots & p_{1n} \\
\vdots & \ddots & \vdots \\
p_{n1} & \cdots & p_{nn}
\end{bmatrix}
\end{equation}
\toolS applies perturbations at two scopes: global, affecting all points in the PCD (100,000-120,000 points), and local, affecting only points in the annotated obstacles in the PCD.
Local points are obtained from KITTI's ground truth describing its location and dimensions. Using the provided calibration files, we project the bounding boxes into 3D space to capture the corresponding obstacle points in the PCD. The point cloud within each bounding box is then extracted, after which perturbations are applied locally by modifying only these points. All other points are masked to ensure that the perturbation effects are isolated to the obstacles.




\subsection{Applying \tool{} perturbations to test trajectory prediction}
\label{method:traj_pred_model}

To connect 3D detection with trajectory prediction, we model extracted detection deviations, such as missed detections or shifts, as perturbations to the trajectory predictor’s input. This allows deviations observed in one dataset to be systematically applied to another, enabling modular evaluation across different dataset–predictor combinations.

In the first step, we extract deviations in the x and y directions from the output of the previous step (Section \ref{sec:local_perturb_method}. Obstacles beyond ten meters are filtered out to focus on those within the trajectory prediction range. We then calculate the deviations at various percentiles, including outliers.
Assuming normal distribution, we use the $Q1$ (25\%) and $Q3$ (75\%) points in both x and y directions as perturbations to the ground truth.
In addition, we partition the outliers into two groups: those above the upper fence ($UF = Q3 + 1.5 * IQR$) and those below the lower fence ($LF = Q1 - 1.5 * IQR$), then select the minimum, maximum, and median from each group.
Thus, we derived eight perturbation values for each direction.

In the second step, we generate input data with the perturbations from the first step. 
As the data contain consecutive frames, we apply perturbations to the historical trajectory of the obstacle using three patterns: \textit{interval}, i.e., once every three frames (1); \textit{all} frames (2); \textit{remove once} (3), where we remove the obstacle from one frame by replacing the position of the obstacle with the values of the previous frame. 
These patterns are designed to be heuristics of observed real availability statistics to reflect common perturbation patterns.

We then run the respective trajectory prediction models on the baseline and each combination of value (step one) and pattern (step two). Afterward, we can compute the point-wise deviation between the new and old trajectories.

\subsection{Applying SORBET to test an industry-grade ADS}
\label{sec:apollo_method}
We envision \toolS as a tool to evaluate the robustness of 3D obstacle detection and the propagation of these results to various ADS components. Hence, it is important to apply \toolS on a complete ADS (e.g., Apollo) in addition to individual DNN models for 3D obstacle detection and trajectory prediction. In this case, the ADS's input is managed by a simulator (e.g., LGSVL) similar to previous works \cite{Tang2022ASO}.

\pa{Testing Apollo's obstacle detection module}. 
Running Apollo 3D obstacle detection is similar to the OpenPCDet's models, bar two steps, i.e., converting the input and the output. To load the data, we convert the perturbed PCD binaries to CSV format to leverage existing testing code in Apollo to load PCDs and run 3D obstacle detection. Using the process described in Figure~\ref{fig:implementation}, we provide Apollo's CNN segmentation model with the original and perturbed PCDs. Once we receive the detected obstacles from post-processing, we convert the irregular shape bounding box to a cuboid to calculate the IoU with the ground truth. From this, we can compute the resulting deviations from the same perturbations similar to other models.

\pa{Testing Apollo's trajectory prediction module.} 
We use LGSVL to provide ground truth data to perception, camera, and traffic light modules in Apollo, which ensures that Apollo's \textit{apollo/perception/obstacles} channel yields 100\% correct obstacles. The simulated scenario consisted of an obstacle car and an Apollo-controlled car running in the same direction for ten seconds in parallel lanes. 

To add perturbations to the prediction module's input, we use Apollo CyberRT~\cite{cyberrt} to create a listener to \textit{apollo$\slash$perception$\slash$obstacles} (1). Then, we redirect the input channel of the prediction module from (1) to \textit{apollo$\slash$perception$\slash$obstacles$\slash$obstacle\_changed} (2). For the baseline, we forward data from (1) to (2) directly. However, when we want to add perturbations, we modify the obstacles from (1) before redirecting them to (2) in real-time to simulate a deviated detection while the location of the original obstacle in LGSVL remains the same.
We set up a new listener for the channel \textit{apollo/prediction} and record the predicted trajectory of each obstacle.
We run simulations on the baseline and perturbations similar to Section~\ref{method:traj_pred_model} and calculate the deviations.

\section{Experiment Setup}
\label{sec:setup}
\pa{Computing Environment.}
We experiment on an Ubuntu 20.04.3 LTS system with an AMD Ryzen 16-core CPU, 256 GiB of memory, and an NVIDIA GeForce RTX 3090. 

\subsection{3D obstacle detection setup for specification-based robustness testing}
\label{sec:setup_3d}

\pa{Datasets.}
We used KITTI 3D Object Detection Evaluation 2017 (KITTI)~\cite{Geiger2012CVPR}, which contains 3,712 training frames and 3,769 test frames. The ground truth contains nearly 20K obstacles, while the evaluated models detect between 9,777 and 13,296 obstacles.
KITTI is an obstacle detection benchmark widely used in literature~\cite{Tang2022ASO}.

We chose KITTI for its high-resolution 64-channel PCD and dense obstacle scenes, which make it well-suited for evaluating detection robustness. Preliminary experiments with Apollo CNN segmentation and PointPillar confirmed its reliability, motivating its usage for the experiment. In contrast, alternatives such as nuScenes provide lower-resolution point clouds with sparser, more distant obstacles \cite{nuscenes, s23229162}, where Apollo and PointPillar showed impractically low detection accuracy, making nuScenes not suitable for our detection experiment.
Thus, we only consider KITTI in the experiment, which we consider a more practical lower bound.

\begin{table}[t]
\centering
  \caption{The 3D obstacle detection models evaluated. 
  }
  \label{tab:models}
  \centering
\resizebox{\columnwidth}{!}{
  \begin{tabular}{lr}
    \toprule
   Model & Type\\
    \midrule
    PointRCNN (2019) \cite{Shi2018PointRCNN3O}& point-based, two-stage \\
    PointPillar (2019) \cite{lang2019pointpillars}            & voxel-based, one-stage \\    
    VoxelRCNN (2020)\cite{deng2021voxel}      & voxel-based, two-stage \\
    TED (2023) \cite{TED}                       & voxel-based, two-stage, depth-based \\
    CNN Segmentation       & CNN segmentation, industry \\
    (Apollo V5.5) \cite{apollo}       & \\

  \bottomrule
\end{tabular}
}
\end{table}


\pa{Subjects.} 
We experiment on five pre-trained 3D obstacle detection models with diverse architectures and sources (Table~\ref{tab:models}). For academic models, from OpenPCDet~\cite{openpcdet2020}, a popular LiDAR-based 3D obstacle detection repository, we selected PointPillar~\cite{lang2019pointpillars}, PointRCNN~\cite{Shi2018PointRCNN3O}, and VoxelRCNN~\cite{deng2021voxel}. In addition, from the KITTI leaderboard, we selected TED~\cite{TED}, one of the top-performing models at the time of our experiments. We chose TED because it is a published work with publicly available code, enabling reproducibility. For an industry-grade ADS, we use the CNN segmentation model from Apollo V5.5, which has been widely evaluated in prior studies~\cite{avfuzz, Tang2022ASO}. Together, these models represent well-known and diverse approaches to 3D obstacle detection.

\pa{Metrics.} Commonly, an obstacle is considered detected if IoU $\geq$ 0.7 for cars and IoU $\geq$ 0.5 for cyclists and pedestrians.
However, in this experiment, to keep track of obstacle deviations regardless of final detection results (to include those that fail to make the IoU thresholds), except for the difference in the number of detected obstacles before and after perturbation (DIFF), we compute the xyz and IoU deviation for obstacles matched to the same ground truth ID as long as the IoU is greater than or equal to 0.25. This threshold is a compromise since lower values show too little overlap for the obstacles to be considered the same.
We summarize the xyz shift via the large deviation count (LDC), which counts the number of obstacles with any dimension deviating more than a defined threshold. We set the threshold at 0.1m, i.e., five times our maximum shift.



\subsection{Trajectory prediction setup for robustness testing against detection-propagated deviations}
\pa{Datasets.} Since trajectory prediction relies on consecutive frames, we used nuScenes~\cite{nuscenes} and LGSVL~\cite{lgsvl} for deviation measurements, as KITTI contains non-consecutive frames.
nuScenes contains 150 scenes of consecutive frames showing the path of obstacles.
LGSVL is a well-known simulator commonly used in previous simulation-based testing for Apollo \cite{Tang2022ASO}. However, \tool{} is not Apollo-LGSVL specific and is extensible for other ADS and simulator combinations.
Since LGSVL's cloud is no longer available, we set up a local server via SORA-SVL-Server~\cite{soralgsvl} to experiment offline.

\pa{Subjects.} We select Trajectron++, a popular trajectory prediction model~\cite{salzmann2021trajectron, Zhang_2022_CVPR, li2021fooling}. Also, we include Apollo's trajectory prediction, which is a part of a production-grade L4 system and has been used since Apollo 3.5.

\pa{Metrics.} The deviation is measured using root mean square error (RMSE), commonly used in prior works~\cite{salzmann2021trajectron, Zhang_2022_CVPR}, where a lower RMSE indicates higher robustness. Two derivations of the RMSE are used: average displacement error (ADE), the RMSE between the original and the deviated predicted trajectory, and final displacement error (FDE), the RMSE between the final predicted position between the baseline and the deviated trajectory. ADE is good for evaluating how well a model performs throughout the entire trajectory, while FDE is critical when the final position matters most (e.g., merging lanes, avoiding obstacles).
Together, they provide a comprehensive view of prediction performance, especially in real-time safety-critical systems like autonomous driving.

\section{Evaluation Results}
\label{sec:results}
\subsection*{RQ1: \RQOne}
\pa{Motivation.}
Even 1-2cm of perturbation may change the detection results of obstacle detection. The specification of popular LiDAR sensors such as Velodyne HDL-64E \cite{Geiger2012CVPR} or OUSTER OS2 \cite{lidar_manuals}, and Velodyne HDL-32E \cite{nuscenes} state that the distance accuracy is 2-10cm.
In the first step of the experiment, we show how these specification-based perturbation changes the detection results.
Then, we investigate the effects of perturbations local to the obstacles, relating to their shape, color, surface diffusion, distance, etc., as described in Section~\ref{sec:perturbations}.

\begin{table*}[t]
\caption{Summary of deviations caused by \tool's perturbations on 3D obstacle detection. 
`Refl.' is short for reflectivity perturbation. 
`Dist.' is short for distance-amplified range inaccuracy. For range inaccuracy, since three distributions are used, we show each metric as \textbf{\underline{U}}niform/\textbf{\underline{N}}ormal/\textbf{\underline{L}}aplacian reflecting the value recorded for perturbations using each distribution.}
\label{tab:rq1_deviations}
\centering
\resizebox{\textwidth}{!}{
\begin{tabular}{l | r | l | r r r | r r | r}
\toprule
    & Baseline  &    & \multicolumn{3}{c|}{Range Inaccuracy}
                                                    & \multicolumn{2}{c|}{Refl.} 
                                                                      & Dist.\\
    & detections  &    & Global (U/N/L)   & Local (U/N/L)    & Directional (U/N/L) & ($\downarrow$) & ($\uparrow$)\\
\midrule 
\multirow{4}{*}{\rotatebox[origin=c]{67.5}{PointPillar}}
    & 13,296 & LDC  & 409/799/768   & 0/409/388   & 489/940/913  & 5,661 & 735  & 1,204\\
    & & \% & 2.3/4.5/4.4&0/2.3/2.2&2.8/5.3/5.2&32.1&4.2&6.8\\
    & & DIFF & 1,278/816/794   & 0/391/421   & 453/279/261  & 2,768 & 537  & 842 \\
    & & \% & 9.6/6.1/6.0 & 2.9/3.2/3.4 & 2.1/2.0/0 & 20.8 & 4.0 & 6.3\\
    \midrule
    \multirow{4}{*}{\rotatebox[origin=c]{67.5}{PointRCNN}}
    & 13,047 & LDC  & 2,297/3,105/3,109 & 17/2,401/2,398 & 2,771/3,428/3,316   & 4,556 & 3,281& 3,623 \\
    & & \% & 14.1/19.1/19.1&0.1/14.8/14.8&17.1/21.1/20.4&28.1&20.2&22.3\\
    & & DIFF & 2,285/1,819/1,786 & 20/1,408/1,292 & 1,521/1,267/1,204   & 3,072 & 1,717& 1,786 \\
    & & \% & 17.5/13.9/13.7& 0.2/10.8/9.9 & 11.7/9.7/9.2&23.5&13.2& 13.7\\
    \midrule
    \multirow{4}{*}{\rotatebox[origin=c]{67.5}{VoxelRCNN}}
    & 11,908 & LDC  & 691/1,082/981   & 0/658/613  & 570/923/891  & 3,910& 984  & 2,025\\
    & & \% & 4.7/7.4/6.7&0/4.5/4.2&3.9/6.3/6.1 &26.8&6.7&13.9\\
    &  & DIFF & 587/762/747   & 0/533/529  & 504/701/688  & 1,786 & 617  & 1,025 \\
    & & \% & 4.9/6.4/6.3  & 0.0/4.5/4.4 & 4.2/5.9/5.8 & 15.0 &5.2&8.6 \\
    \midrule
    \multirow{4}{*}{\rotatebox[origin=c]{67.5}{TED}}
    & 13,064 & LDC  & 0/237/246   & 0/94/95    & 86/99/101 & 1,540 & 104  & 505   \\
    & & \% &  0/1.6/1.7&0/0.6/0.6&0.6/0.7/0.7& 10.4 & 0.7 & 3.4 \\
    &     & DIFF & 0/394/407   & 0/255/246   & 240/260/259  & 1,275 & 236  & 508   \\
    & & \% & 0.0/3.0/3.1 & 0.0/2.0/1.9 & 1.8/2.0/2.0 & 9.8 & 1.8 & 3.9 \\
    \midrule
    \multirow{4}{*}{\rotatebox[origin=c]{67.5}{Apollo}}
     & 9,777 & LDC  & 1,886/1,924/1,864 & 1,570/1,590/1,567 & 1,615/1,621/1,596& 3,431& 1,716 & 1,683 \\
    & & \% & 15.6/16.0/15.5 & 13.0/13.2/13.0 & 13.4/13.4/13.2 & 28.4 & 14.2 & 14.0 \\
    & & DIFF & 638/623/667   & 379/387/372   & 400/403/398  & 1,967 & 402   & 641   \\
    &  & \% & 6.5/6.4/6.8 & 3.9/4.0/3.8 & 4.1/4.1/4.1 & 20.1 & 4.1 & 6.6 \\
\bottomrule
\end{tabular}
}
\end{table*}

\pa{Method.}  
We run OpenPCDet models using their \textit{tools/test.py} script, only modifying the post-detection steps to pass back all detections regardless of the IoU.
To measure deviations from perturbations, we set a baseline for each model in Table \ref{tab:models} by running them on the unaltered KITTI dataset to record the results. Evaluated models detect between 9,777 and 13,296 obstacles (Table~\ref{tab:rq1_deviations}) from almost 20K ground truth obstacles. 

In each iteration, \tool{} generates a perturbed PCD frame from KITTI by applying a single perturbation. This one-at-a-time approach minimizes interference between perturbations and allows for individual evaluation. With 3,769 unique frames in KITTI and 15 distinct perturbations, \tool{} produces a total of 3,769 x 15 perturbed PCD frames for evaluation by the models in Table \ref{tab:models}. For each model and frame combination, \toolS computes the deviation as the absolute difference, i.e., $\hat{x}-x$ for all detected obstacles with their baseline counterparts across all metrics as described in Section \ref{sec:setup_3d}.

\begin{figure*}
\begin{subfigure}{.5\textwidth}
  \centering
  \includegraphics[width=\linewidth]{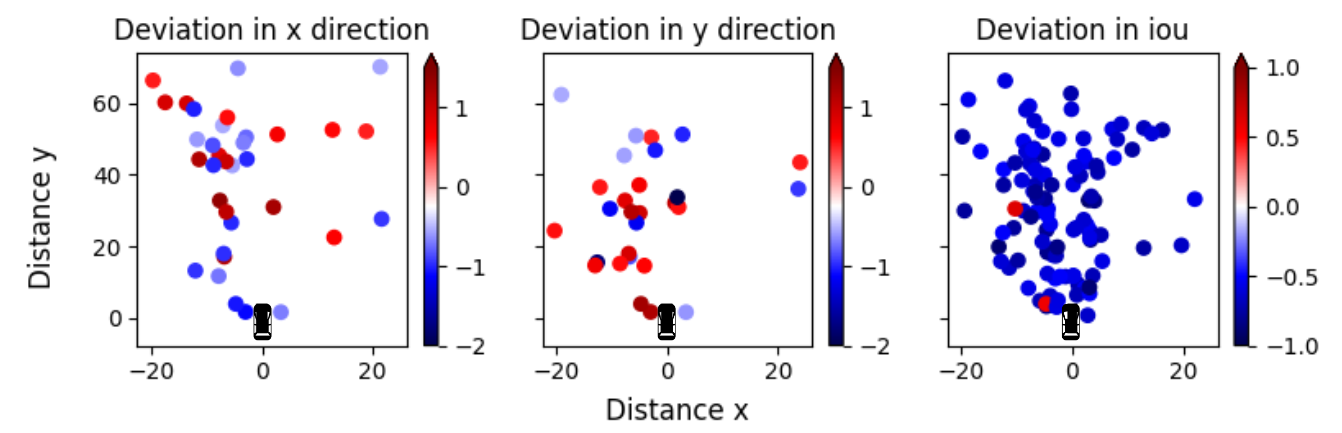}
  \caption{PointPillar}
  \label{fig:fig_global_random_pointpillar}
\end{subfigure}%
\begin{subfigure}{.5\textwidth}
  \centering
  \includegraphics[width=\linewidth]{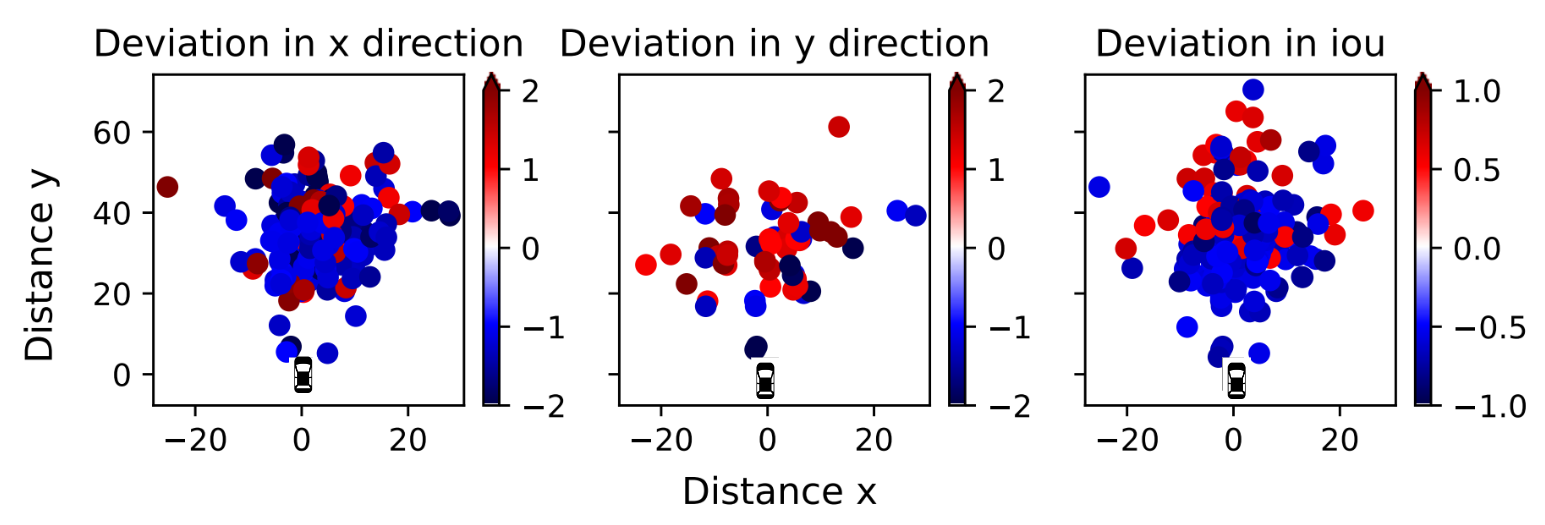}
  \caption{Apollo}
  \label{fig:fig_global_random_apollo}
\end{subfigure}
\caption{Deviations in x, y, and IoU metrics for detected obstacles by (a) PointPillar and (b) Apollo when there is range inaccuracy (global). Only x, y deviations greater than $\pm$0.5 m and IoU deviations greater than $\pm$0.5 (PointPillar) or $\pm$1 (Apollo) are shown. The black car at (0, 0) shows the location of the ego car.}
\label{fig:fig_global_random}
\end{figure*}

\begin{figure}
\centering
\scalebox{1}{
  \includegraphics[width=\linewidth]{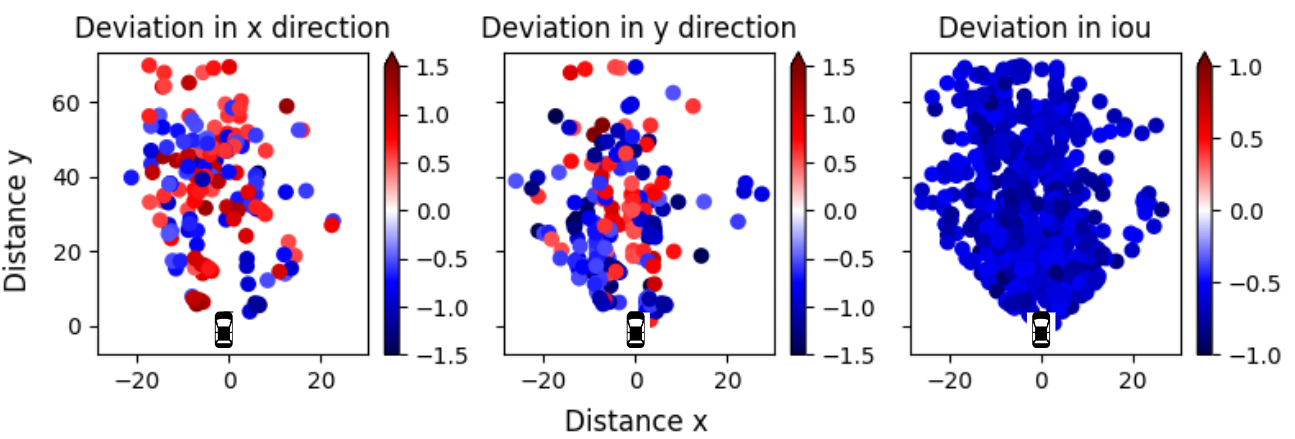}
}
\caption{Deviations in x, y, and IoU metrics for detected obstacles by PointPillar with decreased reflectivity perturbations. Only x, y deviations greater than $\pm$0.5cm and IoU deviations greater than $\pm$0.5 are shown. The black car at (0, 0) shows the location of the ego car.}
\label{fig:fig_local_color}
\end{figure}

\pa{Results.}
On average, the processing time for each frame is 0.03s for PointPillar, PointRCNN, and Voxel-RCNN, 0.09s for TED, and 0.46s for Apollo due to larger overhead.

Our results show an \textbf{amplification effect}, i.e., a 2cm shift in PCD has a median amplification of five to eight times in the dimensions of the detected obstacles, and a maximum deviation of 2.3m for PointPillar.
Overall, we observe that the LDC correlates with DIFF, i.e., an increased number of differences in detection occurs at the same time as the increase in the number of obstacles with large deviations.
Table~\ref{tab:rq1_deviations} shows the robustness results of our study using \tool, showing LDC and DIFF, where snapshots of the results are shown in Figure
\ref{fig:fig_global_random}, \ref{fig:fig_local_color}.

For the \textbf{false positive (FP)} perturbation, Point RCNN and Apollo show non-trivial reduced performance (LDC up to 12.5\% and DIFF up to 2.5\%). Investigation shows that these models are impacted due to their usage of the aggregated confidence score (all points in a detected region) for filtering. The single point removed by FP reduces the aggregated confidence to below a defined threshold, changing the detection results.
This suggests that even changes to \textbf{one point} in the PCD can impact 3D obstacle detection with specific designs.

Our experiments show that \textbf{range inaccuracy (global)} can cause significant deviations (of more than 10cm) in 3D obstacle detection for up to 20\% of the obstacles.
Also, we observe a non-trivial number of obstacles where the IoUs change more than 0.5 (Figure~\ref{fig:fig_global_random}. Considering the standard IoU thresholds, these detections would be considered distinct from the ground truth or previous detection, potentially causing non-trivial downstream issues.
Perturbations of type \textbf{range inaccuracy (local)} result in fewer obstacles having large deviations and smaller decreases in detected obstacles. 
While we observe non-trivial deviations, obstacle-specific perturbations cause fewer undetected obstacles compared to the non-target version. Perturbations generated by uniform randomization even result in zero deviations. 
We observe several contributing factors. 
First, the average IoU for baseline detections is at most 0.8, indicating a large offset from the ground truth, which reduces the effectiveness of obstacle-specific perturbations.
This brings us to the second point that detections are likely impacted by both the points inside and outside of the bounding box; thus, only changing the point inside the ground truth bounding box does not cause the expected impact.
Then, \textbf{range inaccuracy (directional)} causes more obstacles to have large deviations than range inaccuracy (local), even on par with range inaccuracy (global) in some models.
The output confirms our intuition that perturbations applied in the same direction magnify the impact of the perturbation.
This shows that non-trivial deviation in the IoU can be caused by hardly visible perturbations to a few points (less than 1.5\%) in the PCD.

When the \textbf{reflectivity is decreased}, we observe a significant decrease in IoU and number of obstacles detected, i.e., up to 23.5\% compared to baseline, at the same time as a significant increase in the LDC.
The reverse (i.e., \textbf{increased reflectivity}) does not have as much impact but is still on par with range inaccuracy (global) while perturbing significantly fewer points (i.e., 1.5\% of the PCD vs. all). In fact, we observe slightly better detection results compared to the baseline.

Lastly, for \textbf{distance-amplified range inaccuracy}, while the number of obstacles detected is slightly less than range inaccuracy (global), we observe up to 290\% increase in the number of obstacles that have large deviations.

Among evaluated models, TED shows the highest robustness by leveraging additional depth information, while Point RCNN is the most sensitive to point-based perturbations. Apollo's model has moderate robustness among the compared models. 

\rqboxc{RQ1 Finding: While models are robust for the average case, \tool's perturbations have an amplification effect and can cause significant deviations, i.e., up to 17.5\% change in the number of obstacles detected, and up to 20.4\% of detected obstacles show significant deviations in dimensions and size. For local corruption, we observe up to 23.5\% fewer obstacles detected. Failure to detect obstacles compromised the safety and system reliability of autonomous driving systems.}

\subsection*{RQ2: \RQThree}
\pa{Motivation.}
We observe that previous work can improve 3D obstacle detection through retraining with perturbed data or pre-processing to smooth inaccuracies. In the case of TED, the depth is used to enhance the robustness.
While our focus is on testing, it is useful to learn whether these common methods can alleviate the effects of our perturbations. In this experiment, we will attempt retraining and evaluate whether it can improve the models' performance on perturbed PCD.

\pa{Method.}
First, we retrain the models in OpenPCDet that use only PCD with a new training dataset of the same size as the original training data. The frames are sampled from both original and perturbed data so that 33\% of the frames are from the original and 67\% are from the perturbed data. We pick three representative perturbations for this experiment, i.e., range inaccuracy (global, normal), range inaccuracy (local, normal), and reflectivity perturbation (decrease) due to their commonality and large impact.

\pa{Results.}
Table~\ref{tab:rq3_retrain} shows that retraining can improve performance on perturbed data.
We observe better robustness against minor perturbations, reducing DIFF by 0.8\%-10.6\%.
The retrained model shows slightly better performance in perturbed test datasets, i.e., the retrained models detect 0.3\%-0.9\% more ground-truth obstacles.

On the other hand, when evaluated on original data, the performance of the retrained models varies. For range in accuracy (both global and local), we observe that the DIFF and number of detections remain the same with changes at the 0.1-0.2\% range. However, for the reflectivity perturbation, we observe an interesting divergence where DIFF increases by 5.4-6.7\% and the number of detections decreases by 4.4-5.1\%. 
We found that this issue stems from significant changes in the size of the point cloud representations, effectively introducing a new set of obstacle characteristics. As the training dataset size remains fixed, the model has reduced exposure to the original obstacle dimensions.
We will continue the experiment on defense strategies in the next version of the study.

\begin{table}[t]
\caption{Results showing the change in performance between the original and retrained model on perturbed data.}
\label{tab:rq3_retrain}
\centering
\resizebox{\columnwidth}{!}{
\begin{tabular}{l | l  |r|r|r}
\toprule
    &       & RI Global
                & RI Local
                & Refl. $\downarrow$\\
\midrule
PointPillar 
    & Det. (\%)  & 0.9  & 1.1  & 9.9\\
    & DIFF (\%) & -6.7  & -6.8 & -10.1\\
    \midrule
PointRCNN
    & Det. (\%)  & 1.2  & 1.0  & 10.8\\
    & DIFF (\%) & -0.8  & -0.9 & -11.5\\
    \midrule
VoxelRCNN
    & Det. (\%)  & 0.3  & 0.3   & 6.5\\
    & DIFF (\%) & -10.6 & -10.2 & -8.5\\
    \bottomrule
\end{tabular}
}
\end{table}

\rqboxc{RQ2 Finding: Retraining with perturbations is an effective strategy for addressing subtle, specification-based deviations. However, in the case of larger perturbations, retraining introduces a trade-off: while the model becomes more robust to the perturbations, its performance on the original, unperturbed data may degrade.}

\subsection*{RQ3: \RQTwo}
\pa{Motivation.} 
In the previous RQ, we observed that \tool's perturbations can cause significant detection deviations. Thus, we are interested in learning how these deviations propagate, specifically, their impact on trajectory prediction. As undetected obstacles (i.e., DIFF in Table~\ref{tab:rq1_deviations}) clearly reduce ADS safety, our focus is on large deviations in detected obstacles, whose effects are less obvious. Since trajectory prediction relies on an obstacle's historical positions, such deviations may alter the predicted paths. We aim to quantify these trajectory shifts and assess whether they can affect the ego vehicle's behavior, paving the way for future work in planning and control evaluation.

\pa{Method.}
We simulate the baseline and perturbations as described in Section~\ref{method:traj_pred_model} and calculate the deviations.
We evaluate this using two combinations of models and data. Two combinations are used as each has its advantages and disadvantages. Trajectron++ runs on fixed trajectory data from nuScenes and which allows experimentation on a large number of scenarios while producing consistent results across runs for reproducibility. Hence, it is faster to run since we only need to execute it once per perturbation. On the other hand, when we simulate the input for Apollo using LGSVL, it provides more realistic insights into a real system. However, it shows significant fluctuations in the predicted trajectory, making it hard to record consistent output. To enhance reproducibility, we simulate twenty times for the baseline and each perturbation type. This process makes experimenting on a large number of scenarios costly. Combining the two allows a clearer understanding of the effects of propagated deviations on trajectory prediction under different systems with different characteristics while reducing cost.

First, we run Trajectron++ with the 150 scenes in nuScenes.
Second, we run Apollo with LGSVL, where we set up a scenario in the simulation where the obstacle and car run in the same direction for ten seconds in parallel lanes. We record this trajectory as the baseline trajectory. For measurements, we identify the index of the most common frame where the obstacle is detected and its location, as well as the frame's index at which the first prediction is made from detections since the first detection frame. 
This allows us to obtain around eight frames from each perturbation with similar profiles, yielding similar predicted trajectories, where we take the point-wise average of the predictions. The cases in Table~\ref{tab:rq3_results_nuscene_combined} are predicted at the 0.8s point, where the trajectory prediction module in Apollo makes the first prediction after seven to eight frames from obstacle detection. Our approach is a tradeoff where we filter a large number of runs in exchange for more stable results. Hence, we expect to report the most \textbf{reproducible} results, not the highest deviation found.
For Apollo, the input is the simulation ground truth. 
Note that the location of the original obstacle in LGSVL remains the same; we are only adding perturbations based on calculated deviations to the input of the prediction module.


We compute the deviation by comparing the predicted trajectory under perturbed data against that under the original data. Specifically, we run each trajectory prediction model on the baseline input and on every combination of value (step one) and pattern (step two) described in Section~\ref{method:traj_pred_model}. We then compute the ADE and FDE between the original and perturbed trajectories. We compute the ADE and FDE separately for the x and y directions.
To assess the impact of deviated predictions on the ego vehicle, we assume obstacle car dimensions consistent with those in nuScenes~\cite{nuscenes}, which is 1.73m wide. We also assume that the car is at the center of its lane, where the average lane width is 3.6m~\cite{lane_width}, making the distance from the side of the car to the lane line 0.935m.
ADE/FDE exceeding this value indicates a high likelihood of the deviated trajectory's impact on ego' planning.

\pa{Results.}
Since deviation in the x direction is key to unsafe behaviors, we would focus on analyzing this in Table~\ref{tab:rq3_results_nuscene_combined}.
Overall, perturbations in the x direction resulted in smaller deviations than in the y direction.
\begin{table}[t]
  \caption{Top one percent deviations for Trajectron++ (base) on nuScene and Apollo using LGSVL. The deviations are reported as ADE(x)/FDE(x), where the units are in meters.}
  \centering

\resizebox{\columnwidth}{!}{
  \begin{tabular}{ l | rr | r  r | r  r}
    \toprule
              &           &     & \multicolumn{2}{c|}{Trajectron++ (1s)}  & \multicolumn{2}{c}{Apollo (0.8s)}\\
    \midrule
              & \multicolumn{2}{c|}{Perturbations}
                                    & \multicolumn{1}{c}{Interval}    
                                                               & \multicolumn{1}{c|}{All}& \multicolumn{1}{c}{Interval}    
                                                                                             & \multicolumn{1}{c}{All}\\
    \midrule
              & + x      & + y    & \multicolumn{4}{c}{ADE(x)/FDE(x)}\\
  \midrule
    baseline  & + 0      & + 0    & 0.0/0.0   & 0.0/0.0 & 0.0/0.0 & 0.0/0.0\\
    \midrule
    x Q1      & - 0.04  & + 0.01  & 0.09/0.12  & 0.04/0.04  & 0.0/0.0   & 0.21/0.29   \\
    x Q3      & + 0.0   & - 0.02  & 0.04/0.06  & 0.0/0.0    & 0.20/0.25 & 0.16/0.17   \\
    y Q1      & - 0.05  & - 0.03  & 0.13/0.16  & 0.05/0.05  & 0.15/0.22 & 0.21/0.34   \\
    y Q3      & - 0.06  & + 0.02  & 0.11/0.14  & 0.06/0.06  & 0.22/0.27 & 0.79/1.14   \\
    \midrule
    x UF      & + 0.06  & - 0.03  & 0.12/0.16  & 0.06/0.06  & 0.02/0.0  & 0.31/0.40   \\
    x LF      & - 0.10  & - 0.04  & 0.15/0.18  & 0.10/0.10  & 0.34/0.51 & 0.50/0.72   \\
    y UF      & - 0.05  & + 0.08  & 0.10/0.14  & 0.05/0.05  & 0.18/0.27 & 0.74/1.09   \\
    y LF      & - 0.03  & - 0.09  & 0.10/0.14  & 0.03/0.03  & 0.52/0.77 & 0.28/0.40   \\
    x outlier & - 0.34  & - 0.58  & 0.51/0.67  & 0.34/0.34  & 0.0/0.0   & 0.31/0.0    \\
    y outlier & - 0.13  & - 0.14  & 0.19/0.23  & 0.13/0.13  & 0.03/0.01 & 0.22/0.24\\
    \midrule
    x min     & - 5.90  & - 3.16  & 7.04/8.63  & 5.91/5.93  & 2.69/3.50 & 0.01/0.0\\
    x/y max   & + 0.53  & + 0.41  & 0.66/0.80  & 0.53/0.53  & 0.04/0.02 & 0.21/0.23\\
    y min     & - 5.37  & - 3.24  & 6.46/7.97  & 5.38/5.39  & 2.73/3.47 & 0.31/0.51\\
  \bottomrule
\end{tabular}
\label{tab:rq3_results_nuscene_combined}
}
\end{table}

\begin{figure}[htbp]
    \centering
        \centering
        \includegraphics[trim={1cm 1.5cm 2cm 0.5cm},clip,width=0.9\columnwidth]{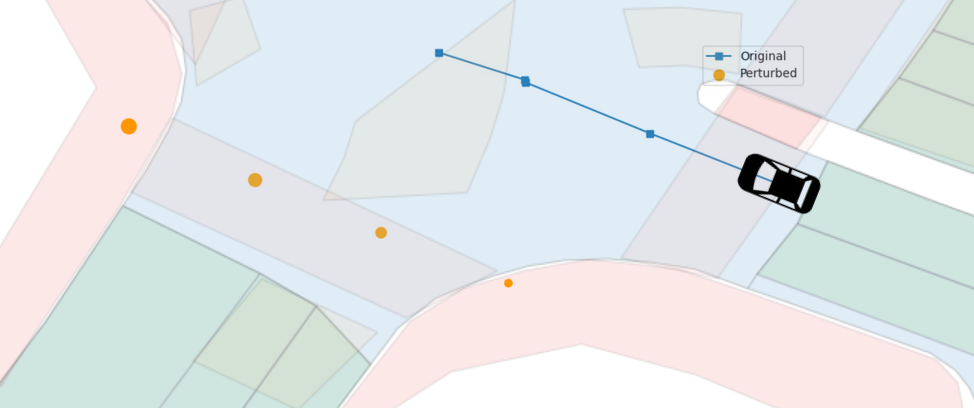} 
     \caption{
Example of an extreme outlier. The obstacle (black car) is at its ground truth position, but its detected position deviates due to perturbation. The minimum x-direction deviation observed from Apollo in RQ1 is applied at an interval as a perturbation to Trajectron++. The resulting predicted trajectory (orange) deviates from the original (blue), shifting onto the sidewalk, with an ADE of 7.04m and an FDE of 8.63m in the x direction. \label{fig:perturbed_trajectory_129}}
\end{figure}

For Trajectron++, the average case for all perturbation patterns shows that the ADE and the FDE are very close to the introduced perturbation, i.e., the largest deviation is less than 0.16m, where the ADE and FDE are mostly identical.

However, the top 1\% of the deviations (by value) show non-trivial deviations (Table~\ref{tab:rq3_results_nuscene_combined}). For the \textit{interval} pattern, the ADE can be up to 3.3 times the original perturbation, and the FDE can even be up to 4.7 times.
For example, in Figure~\ref{fig:perturbed_trajectory_129}, the obstacle on the lane is mispredicted to be on the sidewalk almost 2m away, which will highly likely impact planning.
However, most cases show less than 0.3m in the resulting x deviation, suggesting the obstacles are not predicted to change lanes unless very close to the lane dividers or when the ego is changing into the obstacle's lane.
For the \textit{remove once} pattern, the top 1\% can cause 9.77m of deviation in the x direction, potentially causing hazards.
However, for perturbations using the \textit{all} pattern, the deviations from Trajectron++ are similar to the average case.

On average, Apollo's predicted trajectory deviation, while the deviations are more significant than Trajectron++, is still less than 0.3m.
However, the top 1\% deviations (Table~\ref{tab:rq3_results_nuscene_combined}) are significant and likely to disrupt path planning, where we observed Apollo briefly stop unnecessarily. 
Still, monitoring the planning channel shows that this only has a delay effect and does not change the original planning path.
Surprisingly, unlike Trajectron++, when perturbations are applied to \textit{all} frames, deviations increase significantly, e.g., ADE can increase by up to 17.3 times and FDE by up to 25.6 times.
Hence, perturbations (less than 2cm) can cause deviations large enough to alter predicted obstacle trajectories. 
However, adverse effects are not guaranteed due to frequent changes in prediction and planning decisions. Still, this highlights concerns about obstacle perception in ADS and the need for safeguards to prevent safety risks from momentary errors, e.g., unnecessary stops.

\rqboxc{RQ3 Finding: While trajectory predictors are robust to perturbations on average, less than 2cm specification-based perturbations can lead to substantial prediction errors. Our results show that such small deviations can increase ADE by up to 17.3× and FDE by up to 25.6×, even in an industry-grade autonomous driving system.}

\section{Threats to Validity}
\label{sec:threats}

\pa{Internal Threats.}
We observed that detection results can be unstable for certain models, even for identical inputs, which may reduce confidence in the evaluation. For example, PointRCNN can detect up to seven different obstacles across repeated runs. To ensure reproducibility and improve reliability, we executed each model five times per input and reported the averaged measurements. While these models consistently exhibited small deviations, their results followed the same increasing or decreasing trends observed in more stable models.

\pa{External Threats.}
Among inaccuracies described by specifications \cite{lidar_manuals, Geiger2012CVPR, nuscenes}, which show a range of inaccuracies between 2-10cm. We use the lower number~\cite{Geiger2012CVPR} because if the lower number can cause noticeable deviations, the larger would exacerbate the issue, strengthening our conclusion.


\pa{Construction Threats.}
Our obstacle-specific perturbations are computed using ground truth dimensions for generalizability across models. Although the median IoU between detections and ground truth is typically 70–80\%, suggesting less impact than expected, we still observe non-trivial deviations.

To add perturbations to Apollo's trajectory prediction input, we reroute the data (Section~\ref{sec:apollo_method}), which may risk delays or frame drops. In our experiments, however, the added latency was negligible and did not affect the number or timeliness of predictions.

\section{Discussions and Future Work}
\label{sec:discussion_future}

\pa{Comprehensiveness and Realisticity.}
Unlike prior work that emphasizes synthetic or adversarial corruptions, \tool{} focuses on specification-based perturbations, i.e., errors derived from real-world LiDAR sensor specifications and prior experiments (e.g., distance accuracy, reflectivity, surface diffusion) as described in Section~\ref{sec:perturbations}. These perturbations arise under normal sensor operation rather than during extreme weather or attacks, ensuring realism. Based on available documentation, we believe our list of perturbations captures the key built-in inaccuracies of LiDAR sensors to complement existing works. Nevertheless, the framework is extensible to incorporate additional perturbations as they are identified, which we plan to explore in future work.

\pa{Composite Perturbation Testing for ADS.}
Our preliminary study shows that combining perturbations (e.g., reflectivity and noise) produces greater deviations than individual perturbations. As future work, we plan to develop a systematic approach for multi-source composite perturbation testing, potentially leveraging probabilistic models. In addition, future efforts can explore composite perturbations that integrate extreme conditions (e.g., adverse weather) with our specification-based perturbations, using physics-aware models \cite{Hahner2022LiDARSS} to simulate these effects more realistically.

\pa{Cascading Effects on Planning and Control Modules.}
Overall, deviations in behavior as a result of propagated errors from previous modules remain unexplored \cite{Tang2022ASO}. For immediate future work, we plan to extend the end-to-end safety analysis to the planning module using deviations measured from prediction. The deviations can then be modeled for perturbations to a compatible dataset, e.g., nuPlan, to evaluate the planning deviations. Since planning depends on probabilistic trajectory predictions, we anticipate that the deviation analysis is complex as it must account for multiple predicted paths and varying planning outcomes, which requires dedicated methods and metrics such as those in \cite{pmlr-v229-dauner23a}.

\section{Conclusion}
\label{sec:conclusion}
Our study highlights the importance of evaluating the robustness of 3D obstacle detection against specification-derived perturbations and demonstrates how such evaluations can guide retraining to improve model resilience and performance. The resulting output deviations induced by these perturbations are shown to propagate to downstream trajectory prediction, revealing system-wide performance degradation. Our results indicate differing levels of robustness across detection architectures and configurations, which should inform model selection and optimization. Consequently, developers of ADS 3D obstacle detection modules should incorporate robustness testing against LiDAR specification-derived perturbations to ensure comprehensive and realistic model evaluation.
\tool's implementation is available at~\cite{lidar_manuals}.

\bibliographystyle{IEEEtran}
\bibliography{paper}

\end{document}